# A new protocol implementing authentication transformations for multi-located parties

Pradeep Kumar Dantala
Computer Science Department
Oklahoma State University, Stillwater, OK 74078
pradeep.dantala@okstate.edu

**Abstract:** This paper discusses a new protocol implementing authentication in a multi-located environment that avoids man-in-the-middle (MIM) attack, replay attack and provides privacy, integrity of a message for multi-located parties. The protocol uses the concept that each party is associated with a subsidiary agent.

## 1 Introduction

Classical cryptography uses either public-private key pair or single secret shared key for encryption and decryption of a message. In a communication between two nodes, there is always a chance of MIM attack [1]. In the new encryption scheme for multi located parties by Kak [2], it is impossible for a localized eavesdropper to get control of the entire conversation since the points of entry and exit of data can be far apart physically. This note describes a modification to Kak's three stage protocol for multi-located parties, which provides greater security.

Basically, in distributed cryptography, the objective is to share the secret among several parties, similar to a case in a bank where $k$ out of $n$ officers use keys simultaneously to open a vault [3],[4]. Here each party is supposed to have computers at different locations and communication among them is secure. In the simplest case, we consider each party to have one main location and a subsidiary location that we call its agent. The links between agents of different parties are not secure.

In standard cryptography we use single transformations on data whereas in multi located parties we use transformations in sequence by several parties that guarantees the authentication. Furthermore, encrypted data can be divided into several modules or portions and sent over different channels. Multiple paths between sender and receiver make it easier to implement joint encryption and error correction coding, which cannot be achieved in such simplicity in traditional cryptography [5],[6],[7],[8].

In the case of multi located parties, the task of the eavesdropper is complicated by the fact that there exist multiple paths for the sender to send the information to the receiver. This makes it possible to develop rich protocols to guarantee security that take advantage of the multiple paths between sender and receiver. This will be done using the Needham-Schroeder symmetric key protocol for classical authentication [9]. There exists some overlap between these ideas and that of secret hardware, public key cryptography [10].

In this paper, we first discuss different kinds of grammars to convert a given message into production sequence numbers and how these grammars can be applied to linguistic transformations. After that we will discuss systems based on multi-located parties and Kak's three stage protocol for multi-located parties. The last section of the paper describes a new modified protocol implementing authentication and linguistic transformations for multi-located parties.

## 2 Background

We begin with a discussion of how the message is converted into a string of symbols using the production rules of an appropriate grammar G, which is a quadruple (V, T, P, S) defined as follows:

V- Finite set of Non-Terminals or variables
T-Finite set of Terminals
P-Finite set of productions or rules defining a grammar
S-Distinguished non-terminal called as start symbol

Example of simple grammar:
G= ({A, B, S}, {0, 1}, {S→AB, A→0B, B→10A, A→}, {S})

Grammars may be divided into four classes by gradually increasing restrictions on the form of productions in the Chomsky hierarchy.

In this paper we use linguistic transformations based on context free grammars. The objective of splitting and sharing secret information is to generate the data in secret that can be shared by multiple authorized parties [11]. The general methodology using grammars for secret sharing among multiple parties consists of several steps such as

1. Select a classical scheme for secret sharing
2. Convert source data in the form of bit sequences
3. Define grammar for generating secret for input message
4. Using syntax analyzer to parse the bit sequence
5. Generate sequence of grammatical rules
6. Split the secret with selected threshold scheme
7. Distribute the secret among multiple parties of the protocol

Sender A wants to transmit a message to B. Sender A converts message into binary bits and selects an appropriate grammar to convert bits into production sequence number and B, who already know the grammar, can obtain the secret after receiving the production sequence number.

# 3 System Based on Multi-located Parties

The system consists of sender A and receiver B who are globally distributed. Sender A has agent near B's location and receiver B has agent near A's location. Figure 1 shows secure communication for multi-located parties.

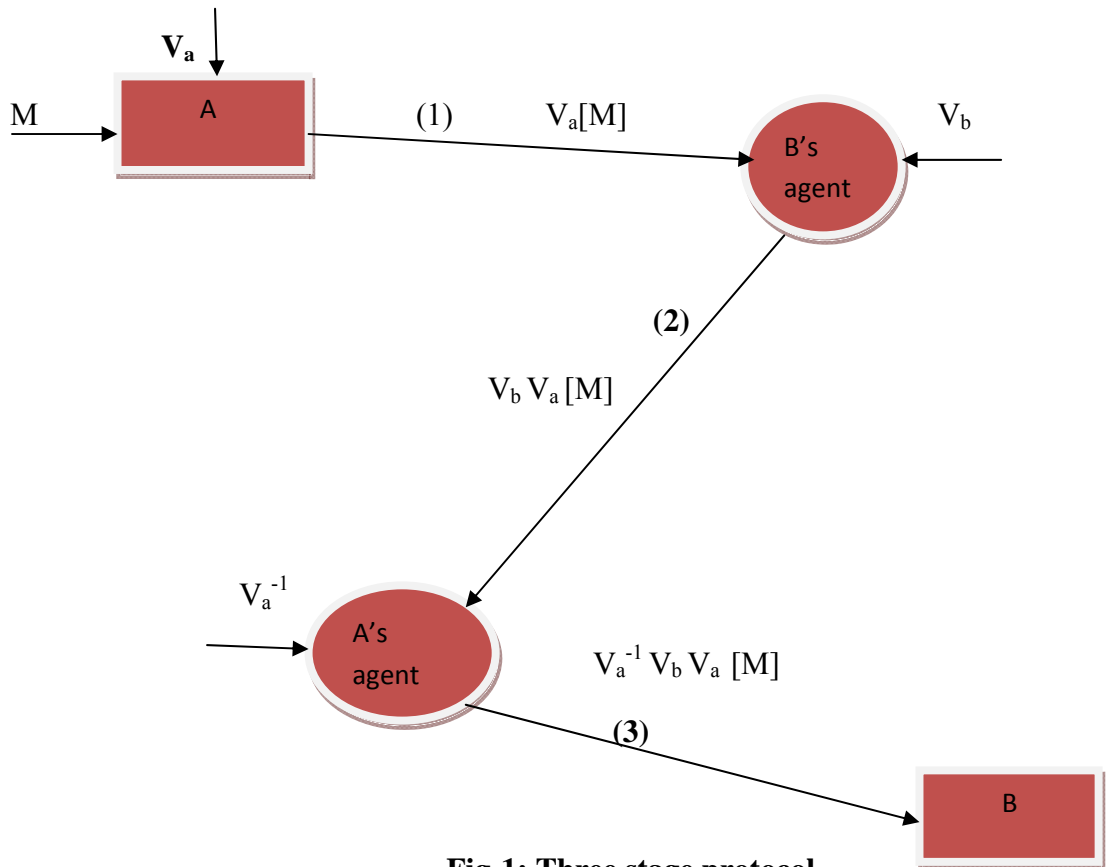

**Fig-1: Three stage protocol**

**Observations:**

| A | A's agent | B | B's agent |
|---|---|---|---|
| A ($V_a$) | A's agent ($V_a^{-1}$) | B ($V_b^{-1}$) | B's agent ($V_b$) |

**Protocol:**

1) A→ B's agent : $V_A[M]$

2) B's agent → A's agent : $V_B V_A[M]$

3) A's agent → B : $V_A^{-1} V_B V_A[M]$

4) B finally performs transformation such that $V_A V_B = V_B V_A$

In the above case, both nodes A and B do not trust each other so they use secret transformations $V_a$ and $V_b$ for authenticating transmissions between nodes in such a way that $V_a V_b = V_b V_a$. In this protocol, A performs transformation on message M and sends it to B's agent; B's agent also perform transformation on received message and send it to A's agent. A's agent will perform inverse transformation on received message and transfer it to B. finally B applies $V_b$ on received message to get the original message. In this way data is transformed securely avoiding main in the middle attack.

**3.1 Three stage protocol for multi-located parties**

We now outline a modification to the above scheme where the information available to the parties and their agents is not identical.

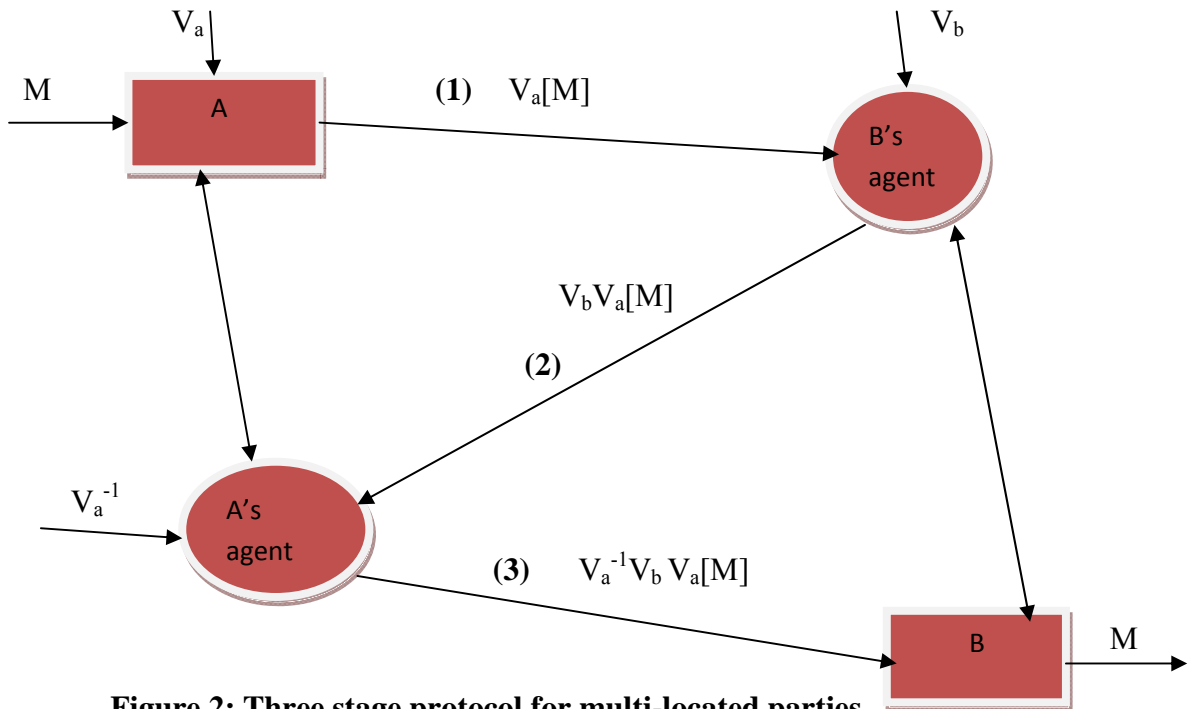

**Figure 2: Three stage protocol for multi-located parties**

This case can be viewed as, nodes A and B are distributed and it uses multiple agents for communication between A and B. The double arrow mark in the Figure 2 represents secure communication channel.

Observations:

| A | A's agent | B | B's agent |
|---|---|---|---|
| A ($V_a$, $V_a^{-1}$) | A's agent ($V_a^{-1}$) | B ($V_b$, $V_b^{-1}$) | B's agent ($V_b$) |

## 4 A new modified three stage protocol for multi-located parties

This section discusses the proposal of new modified three stage protocol implementing authentication based on Needham-Schroeder protocol and linguistic transformation using grammars for multi-located parties.

**Description of protocol**

1.
    Node A → BB: $V_{AB}$

    Node A → AA: $V_{AA}$

    Node A → KDC: E ($K_A$, [$ID_A$, $ID_B$, $N_a$])

2.
    BB → KDC: E ($K_B$, [$ID_B$, $N_b$, $T_b$, $V_{AB}$])

3.
    KDC → AA: E ($K_A$, [$ID_B$, $N_a$, $T_b$, $K_s$]) || E ($K_B$, [$ID_A$, Ks, $V_{AB}$])

4.
    AA → Node B: E ($K_B$, [$ID_A$, Ks, $V_{AB}$]) || E (Ks, [Nb, $V_{AA}$])

5.
    Node B: ($V_{AA} V_{AB}$, G) → Input message

Figure 3 depicts the following steps for multi-located parties:

   1)
       Node A perform transformations using grammar and divide it into two parts $V_{AA}$, $V_{AB}$ and send $V_{AA}$ to AA and other to BB and encrypted information of ID's of

nodes A & B with a nonce Na to KDC using shared key $K_A$ between node A and KDC.

2) BB sends encrypted form of information containing his ID, nonce Nb and time stamp Tb along with $V_{AB}$ to KDC using shared key $K_B$ between BB and KDC.

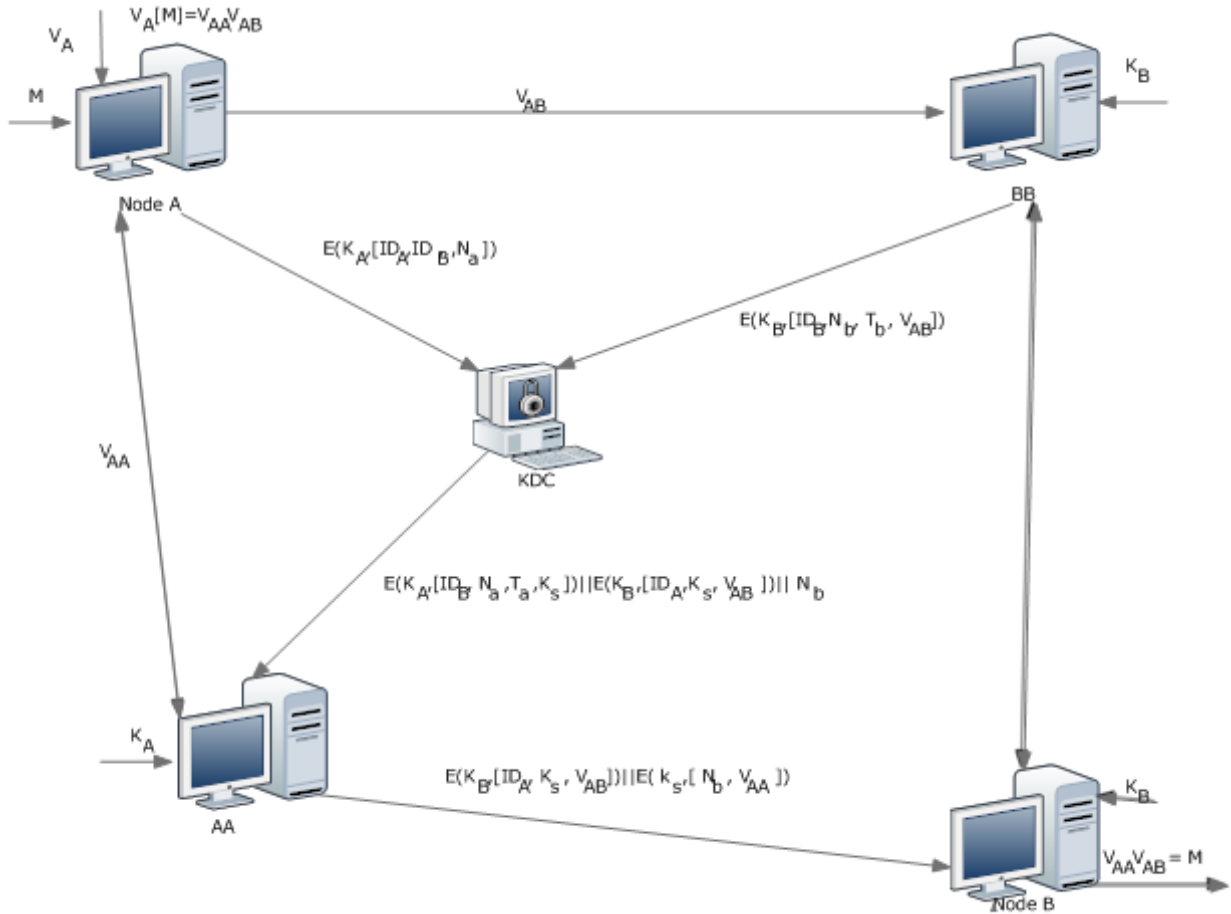

**Figure 3: Protocol implementing authentication and linguistic transformations for multi-located parties**

3) KDC assigns a session key and send information to AA containing identity of B, nonce Na, time stamp Tb and session key Ks which is encrypted using A's key $K_A$ and information containing identity of A, a session key and $V_{AB}$ which is encrypted using B's key $K_B$.

4) AA receives his nonce Na back and A is assured of timeliness by the session key and ensured that it's not a replay. AA send information containing ID of A, session key

Ks and $V_{AB}$ which is encrypted using B's key $K_B$ and encrypted form of nonce Nb and $V_{AA}$ using session key Ks.

5) Finally, B receives two parts $V_{AA}$ and $V_{AB}$ and combines both. B use $V_{AA}V_{BB}$ and grammar to reveal the secret.

Observations:

| A | AA | B | BB |
| --- | --- | --- | --- |
| A ($V_A[M]$, $K_A$, grammar, bit sequence number) | AA ($K_A$) | B ($K_B$, grammar, bit sequence number) | BB ($K_B$) |

Sender A performs linguistic transformations using grammars to convert a message into production sequence numbers. Destination B uses production sequence numbers and grammar to reveal the secret.

---

Example: Linguistic transformations using grammars on input message

Input Message: "hello"
↓
Converting message into bits using 7 bit sequence:10000111010011001101111111011
↓
Sender A- select bit sequence(e.g. 3 bit sequence): <u>100</u>00<u>111</u>0<u>100</u><u>110</u>0<u>110</u><u>111</u>111<u>011</u>
↓
Sender A-select any type of grammar(e.g. context free grammar using 3 bit sequence)
G=({S,A},{0,1},P, {S})
Where P is defined as follows
1 S→BB
2 B→AB
3 B→€
4 A→000
5 A→001
6 A→010
7 A→011
8 A→100
9 A→101
10 A→110
11 A→111
↓
Use grammar to convert bits into production sequence number:1282521028272112=$V_A[M]$

---

In Figure 4, sender A performs linguistic transformations on input message to produce $V_A[M]$. Sender A will break transformed message into two parts $V_{AA}$, $V_{AB}$ and transmit it to two agents AA and BB. BB encrypts message $V_{AB}$ and send it to AA, who in turn send both $V_{AA}$, $V_{AB}$ to destination B.

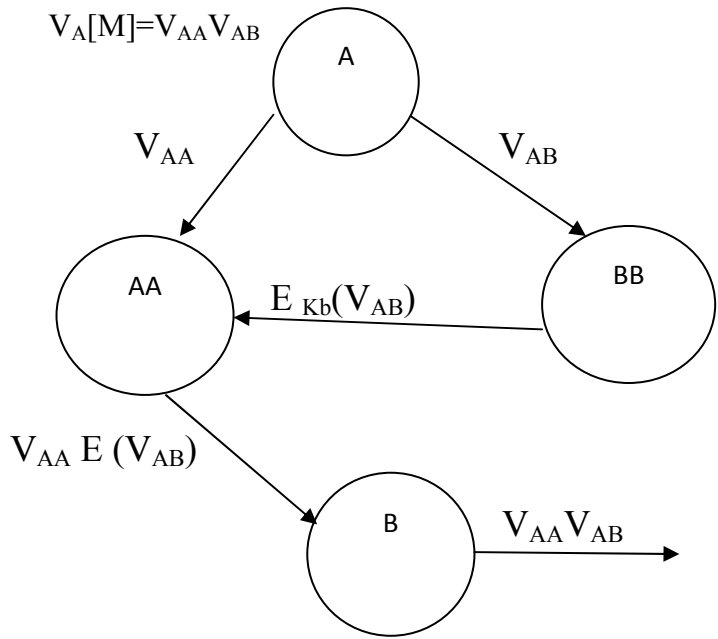

**Figure 4. Protocol for linguistic transformations and secret sharing for multi-located parties**

## 5 Analysis of proposed protocol

For two parties, A and B, that are globally distributed and want to exchange information among them, there are two other parties namely AA for A's agent and BB for B's agent, who participate to securely communicate information from source to destination. As discussed in the previous section, presence of KDC allows all the parties to authenticate and validate themselves before transmission. Since sender A performs transformations on input message using grammars, we will get a message that represents production sequence numbers. The most important part of the proposed protocol is that A breaks the transformed message into two parts and transmits these two parts to AA and BB.

Since AA or BB know neither the grammar nor complete transformed message, they cannot construct the original message and also it is difficult for intruders to obtain the secret since transformed message is divided into parts and transmitted separately, avoiding MIM attack.

Finally, rather than basing security on direct communication between two parties, the proposed protocol uses the concept of multiple parties between source and destination making it possible to obtain higher level of security. When two parties are globally distributed, this protocol can be implemented in a simple and secure manner.

## 6 Weaknesses in the proposed protocol

The proposed protocol may suffer from denial of service attack [12]. Since it is a distributed network and requires multiple agents to transmit information from source to destination, the attacker may overload the traffic to particular agent or multiple agents in a network so that network becomes slow and some agents are temporarily unavailable.

Although the attacker may not get valuable or secret information from the network, he can slow down or stop the communication between multiple agents resulting in failure of transmitting the information completely or in time.

## 7 Conclusions

This paper presents a new modified protocol for multi-located parties. It implements both authentication and linguistic transformations in a multi-located environment. It solves the problem of MIM attack and replay attack for multi-located parties. The protocol may be strengthened by including the idea of puzzles [13] and that of recursive hidden secrets [14],[15].

## References

1. B. Schneier, Applied Cryptography, John Wiley, New York, 1996.

2. S. Kak, "Cryptography for multi located parties." Cryptography and Security, 2009. arXiv:0905.2977v1.

3. P. Rogaway and M. Bellare, Robust computational secret sharing and a unified account of classical secret-sharing goals. ACM Conference on Computer and Communications Security, pp 172–184, 2007.

4. V. Vinod, A. Narayanan, K. Srinathan, C. P. Rangan, and K. Kim, On the power of computational secret sharing. Indocrypt *2003*, vol. 2904, pp. 265–293, 2003.

5. S. Kak and A. Chatterjee, On decimal sequences. IEEE Transactions on Information Theory, IT-27: 647 – 652, 1981.

6. S. Kak, Encryption and error-correction coding using D sequences. IEEE Transactions on Computers, vol. C-34, pp. 803-809, 1985.

7. S. Kak, New results on d-sequences. Electronics Letters, 23: 617, 1987.

8. A. Parakh and S. Kak, Online data storage using implicit security. Information Sciences, vol. 179, pp. 3323-3331, 2009.

9. R. Needham and M. Schroeder, Authentication Revisited. Operating Systems Review." January 1987.

10. S. Kak, On secret hardware, public-key cryptography. Computers and Digital Technique (Proc. IEE - Part E), vol. 133, pp. 94-96, 1986 .

11. Marek R. Ogiela, Urszula Ogiela, Linguistic cryptography threshold schemes. International Journal of Future Generation Communication and Networking Vol. 2, No. 1, March, 2009

12. K. Park and H. Lee. "On the effectiveness of probabilistic packet marking for IP traceback under denial of service attack." Tech. Rep. CSD-00-013. Purdue University. June 2000.

13. S. Kak, On the method of puzzles for key distribution. Intl. Journal of Computer and Information Science, vol. 14, pp. 103-109, 1984.

14. M. Gnanaguruparan and S. Kak, Recursive hiding of secrets in visual cryptography, *Cryptologia*, vol. 26, pp. 68–76, 2002.

15. A. Parakh and S. Kak, A recursive threshold visual cryptography scheme, *Cryptology ePrint Archive, Report 535*, 2008.